\documentclass{elsart}

\usepackage{graphicx}

\usepackage{amssymb}

\begin{document}

\begin{frontmatter}



\title{The merging cluster Abell 85 caught between meals by XMM-Newton}


\author{Florence Durret}
\address{IAP, 98bis Bd Arago, 75014 Paris, France}
\author{Gast\~ao B. Lima Neto}
\address{IAG/USP, R. do Mat\~ao 1226, 
05508-090 S\~ao Paulo/SP, Brazil}
\author{William Forman}
\address{Harvard Smithsonian Center for Astrophysics, 
60 Garden St, Cambridge MA 02138, USA}

\begin{abstract}
Our XMM-Newton observations of Abell 85 confirm the extended 4~Mpc
filament first detected with ROSAT, which has an X-ray temperature of
about $\sim 2\,$keV and is probably made of groups falling on to the
cluster. A comparison of the temperature map with numerical
simulations show that Abell~85 had intense merging activity in the
past and is not fully relaxed, even in the central region.  Finally, a
deprojected temperature profile has been calculated and used, together
with the suface brightness, to estimate the entropy and pressure
profiles. Abell~85 only presents a mild flattening of the entropy
profile in the center, showing no evidence of an ``entropy floor''.
\end{abstract}

\end{frontmatter}




\section{Introduction}

Abell~85 is a very well studied structure both at X-ray [with ROSAT
(Pislar et al. 1997, Lima Neto et al. 1997), and with BeppoSAX (Lima
Neto et al. 2001)], and optical wavelengths, with extensive redshift
and imaging catalogues (Durret et al. 1998a, Slezak et al. 1998); the
mean galaxy redshift of Abell~85 is $z=0.0555$ (Durret et al. 1998b)
(at $z=0.0555$, 1 arcmin $= 90.5 h_{50}^{-1}\,$kpc, assuming
$\Omega_M=0.3$, $\Omega_\Lambda=0.7$). It comprises a main cluster, a
south blob (a group at the same redshift), a brighter zone south-west
of the cluster center coinciding with a Very Steep Spectrum Radio Source
relic (hereafter the VSSRS) and an extended filament at least 4~Mpc
long. This filament was first discovered with ROSAT by Durret et
al. (1998b) and confirmed by XMM-Newton (Durret et al.  2003); it is
likely to be a chain of several groups of galaxies, falling on to the
main cluster. A comparison of the images of this filament obtained
with both satellites is displayed in Fig.~1.  The spectral analysis of
the filament gives an X-ray temperature between 1.5 and 2.8~keV,
implying that it is probably made of groups falling on to the
cluster. This interpretation is strengthened by the fact that the gas
in the impact region, located between the cluster center and the south
blob is indeed hotter, as expected if it was shocked by the infall of
groups on to the cluster.

\begin{figure*}[ht]
\centering  
\includegraphics[width=13truecm]{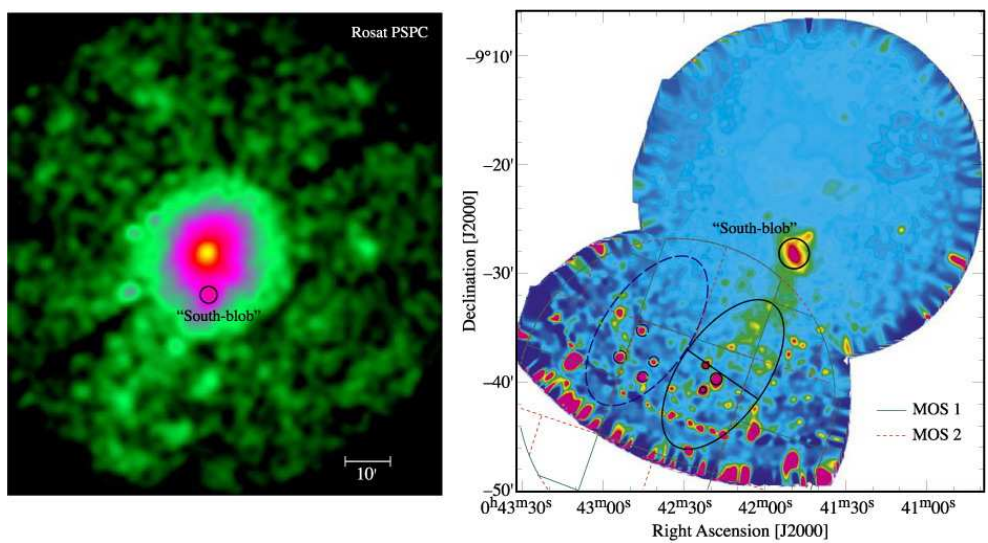}
\caption[]{Left: ROSAT image of Abell~85 and its X-ray filament
towards the south-east.  Right: Merged XMM MOS1 and MOS2 image of
Abell 85, obtained after subtracting an azimuthal average of the X-ray
emission of the overall cluster. The bright diffuse source corresponds
to the south blob (black circle).  The ellipse covering the filament
shows the region in which the spectral analysis was done; the dashed
ellipse is the region where the cluster contribution was estimated
(see Durret et al. 2003).  Note that these two images are not exactly
at the same scale.}
\label{fig:Xray-all}
\end{figure*}

Detailed results on the XMM-Newton analysis of the overall cluster can
be found in Durret et al. (2005); they are complementary to those of
Kempner et al. (2002) who analyzed Chandra data with higher spatial
resolution but in a smaller field. We present here several new results
that complete our study of Abell~85 in X-rays.

A full description of the observations and data reduction can be found
in Durret et al. (2005) and will not be repeated here.

\section{Global properties of the X-ray gas}

A general map of Abell~85 in X-rays showing in particular the VSSRS,
the impact region and the South Blob can be found in Durret et
al. (2005).

\begin{figure}[!h]
\centering
\includegraphics[width=\columnwidth]{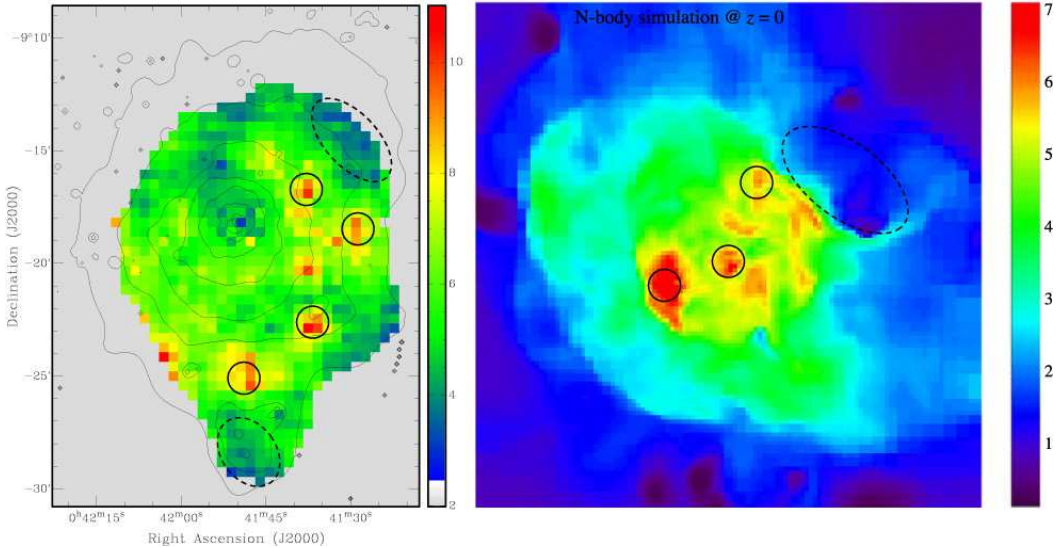}
\caption[]{Left: temperature map and XMM isocontours (clearly showing
the position of the south blob), in scale of keV. Right: result of the
numerical simulation of a cluster merger by Bourdin et al. (2004),
where the merging cluster is assumed to have arrived from the
north-west (dashed ellipse towards the NW) and has caused several
regions to be hotter (black circles). On the left figure, the second
dashed ellipse coinciding with the south blob shows where the second
merging event has originated. A temperature scale in keV is displayed
on the extreme right of the figure.}
\label{fig:tx_z}
\end{figure}

We have derived overall temperature and metallicity maps in a grid,
where each pixel is $512 \times 512$ XMM EPIC physical pixels, i.e.,
each cell grid is $25.6'' \times 25.6''$. In each of these large
pixels we make a spectral fit to determine simultaneously the
temperature and metallicity.

The temperature map is shown in Fig.~\ref{fig:tx_z}. It shows several
interesting features: 1)~it is colder near the center; 2)~two zones,
located at the northwest and southeast of the cluster center, are
respectively cooler and hotter than the average cluster; 3)~the
``impact'' region where the groups forming the filament are thought to
hit the cluster (just north of the south blob) is hotter, as expected
if there is a shock; 4)~the VSSRS is somewhat cooler; 5)~several
hotter patches are observed on the west half of the cluster. Except in
the northeast region where the metallicity is higher but the
temperature shows no variation relative to the average cluster
temperature, the metallicity is anti-correlated with the X-ray gas
temperature (see metallicity map in Durret et al. 2005).  The right
part of Fig.~\ref{fig:tx_z} shows a temperature map resulting from the
numerical simulation of a cluster merger by Bourdin et al. (2004);
this simulation assumes that a seconday cluster falls from the
north-west on to the main cluster. The resemblance between both
temperature maps is striking: several zones of higher temperature are
clearly seen in the simulation, comparable to those observed in the
temperature map.

The combination of the temperature and metallicity maps, and the
comparison with the temperature map derived from the numerical
simulation of Bourdin et al. (2004) strongly suggest that Abell~85 has
undergone several mergers.  As mentioned above, the ``filament''
southeast of the cluster is probably made of groups falling on to the
main cluster and hitting it in the Impact region. Besides, there has
probably been at least one other rather old merger (older than 4~Gyr),
from the NW towards the SE. The absence of an intensity enhancement in
any of the hotter or more metal rich regions suggests that there is no
gas density discontinuity.

\begin{figure}[!ht]
\centering
\includegraphics[width=8.6cm]{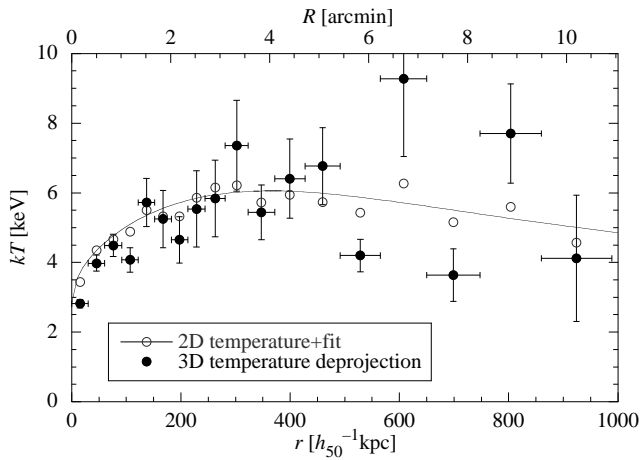}
\caption[]{Gas temperature profile obtained with XMM-Newton without
deprojection (filled circles) and with deprojection (empty squares). The 
curve corresponds to the empirical fit to the 2D temperature profile by 
Durret et al. (2005).}
\label{fig:prof_T_Z}
\end{figure}

The gas temperature profile was derived as well as the deprojected
temperature profile. The 3D temperature profile was obtained following
the method described by Churazov et al. (2003), a non-parametric
method (assuming spherical symmetry) formulated as a least-squares
minimization problem.

As seen from Fig.~\ref{fig:prof_T_Z}, the projected gas temperature
profile shows an obvious decrease towards the center for radii below
about 2.5~arcmin (250~kpc), a roughly constant value between radii of
250 and 450~kpc.  Beyond 450 kpc, the best fit temperature values show
a slow decrease, but the uncertainties are large and the profile is
also consistent with a constant value. A decreasing profile agrees
with that found e.g. by Markevitch et al. (1999).  These variations
were taken into account when deriving the dynamical mass. The
deprojected temperature profile shows a comparable behaviour, but with
much larger error bars.

\section{Gas entropy and pressure}

\begin{figure}[!ht]
\hbox{\kern-0.6cm\includegraphics[width=7.4cm]{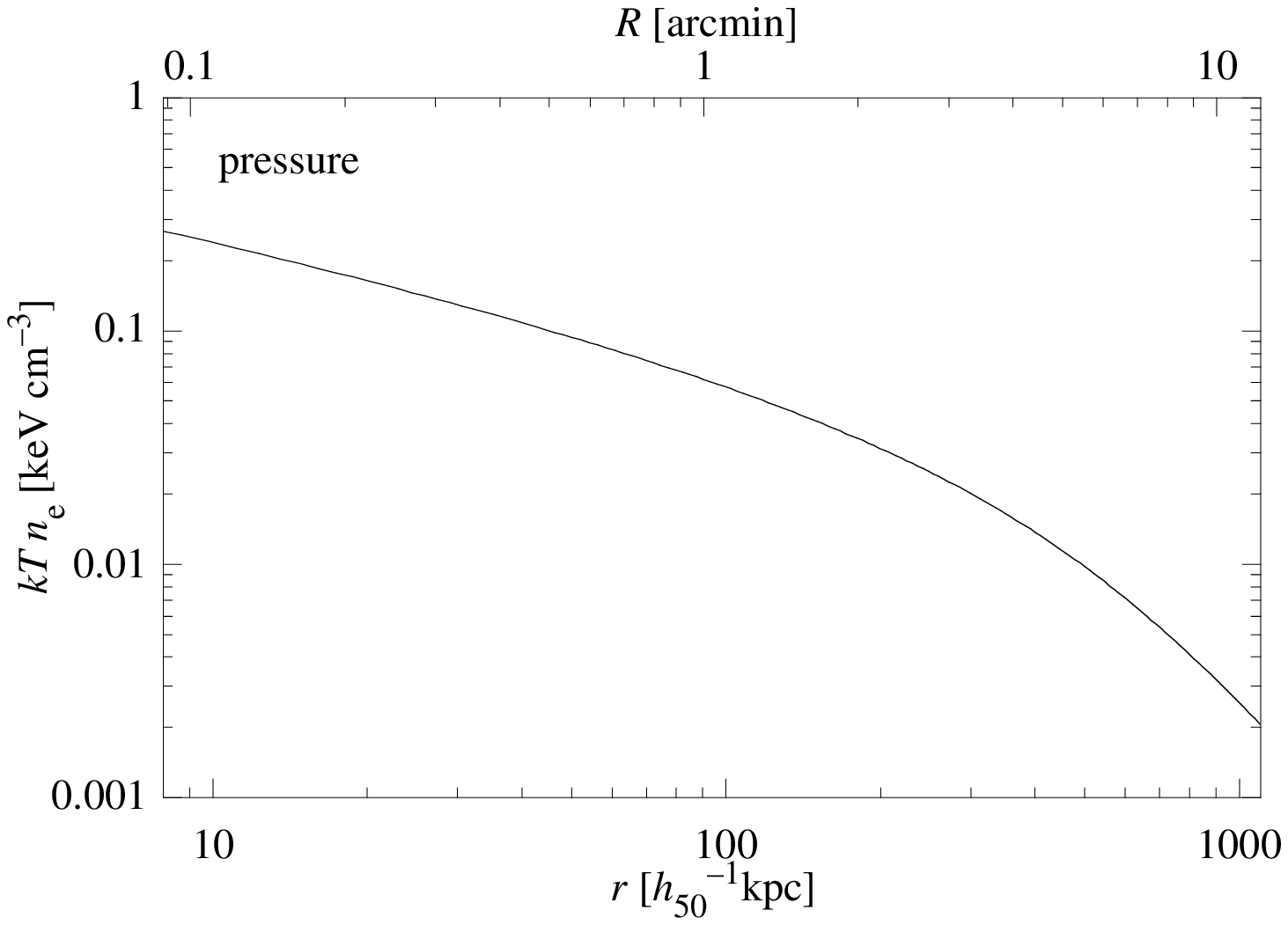}
\kern-0.2cm\includegraphics[width=7.4cm]{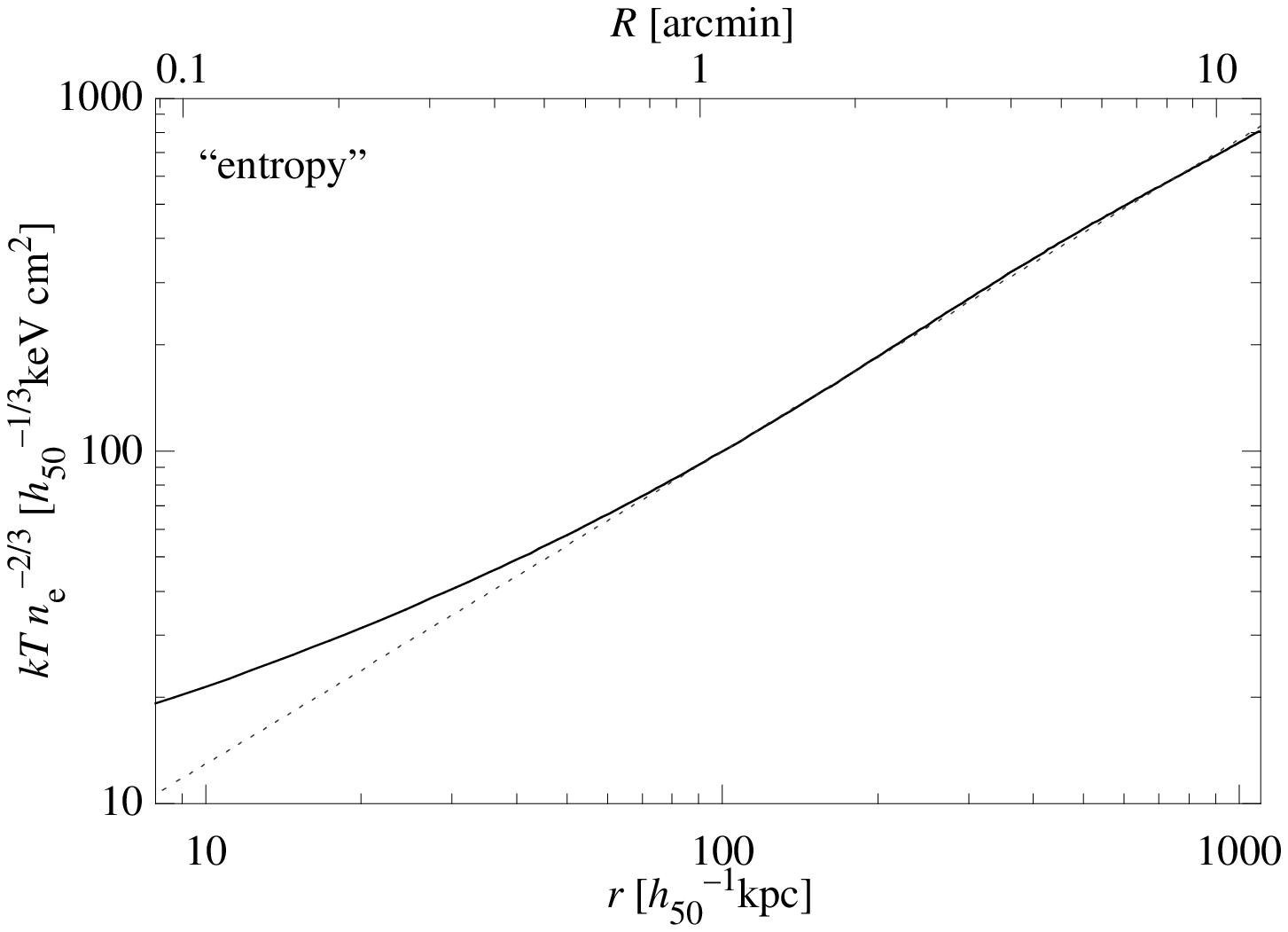}}
\caption[]{Left: gas pressure profile. Right: gas ``entropy'' profile;
the dotted line corresponds to a self-similar entropy profile. }
\label{fig:press_entropie}
\end{figure}

Figure~\ref{fig:press_entropie}, left panel, shows the gas pressure profile. 
Even at our external limiting radius of $1\, h_{50}^{-1}$Mpc, we note that the 
pressure is at least an order of magnitude higher than in gas rich galaxies. 
The gas rich galaxies that fall onto Abell~85 should not be able to keep 
their gas and may lose it before arriving at the core. Thus, in the centre, 
the metal enrichement would be predominantly due to SN~Ia from old stellar 
generations.

The gas ``entropy'' profile is shown in Fig.~\ref{fig:press_entropie}, right
panel. We call ``entropy'', following the trend in cluster litterature, $S
\equiv kT/n^{2/3}$, which is related to the true measure of the gas specific
entropy, e.g., Metzler \& Evrard (1994). The entropy profile shown here was
derived from the surface brightness and temperature analytical fits, cf.
Durret et al. (2005).

The bulk of the cluster shows a power-law entropy profile, with $S
\propto R^{0.9}$, close to the relation $S \propto R^{0.95}$ found by
Piffaretti et al. (2005) for a sample of 13 ``cool-core'' clusters.
The profile slightly flattens near the centre; however, down to the
limiting resolution ($\sim 8\,$arcsec) there is no indication of an
entropy floor, as seen in some groups and clusters (e.g., Lloyd-Davies
et al. 2000; McCarthy et al. 2003), reaching a value as low as
$20\,h_{50}^{-1/3}$~keV~cm$^{2}$ at $R = 10
h_{50}^{-1}$kpc. Nevertheless, the entropy value estimated at $R =
0.1\, R_{\rm virial}$ ($R_{\rm virial} = 2 h_{50}^{-1}\,$Mpc for Abell
85, cf.  Durret et al. 2005) is $S \approx
200\,h_{50}^{-1/3}$keV~cm$^{2}$, comparable to the values derived by,
e.g., Lloyd-Davis et al. (2000).

The low entropy observed in the core may indicate that the ICM has
only a mild heating source in the centre, as expected from the fact
that its temperature profile decreases towards the centre but not
below $\sim 2\,$keV.  This implies that no recent merger has strongly
affected the gas temperature at the center, and therefore that mergers
have not had any strong effect on the entropy profile in the central
regions.

\section{Conclusions}

The interpretation of the temperature and metallicity maps derived for
the X-ray gas in Abell~85 based on numerical simulations by Fujita et
al. (1999) and Bourdin et al. (2004) leads us to suggest that Abell~85
has undergone at least two mergers. First, the ``filament'' southeast
of the cluster is probably made of groups falling on to the main
cluster and hitting it in the Impact region (Durret et
al. 2003). There has most probably also been a rather old merger
(older than 4~Gyr), from the NW towards the SE.  However, the absence
of an intensity enhancement in any of the hotter or more metal rich
regions suggests that there is no gas density discontinuity.  We
intend to achieve more detailed modeling through optical imaging
shortly.

\end{document}